\documentstyle[12pt]{article}
\begin{document}
\setlength{\baselineskip}{3.0ex}

\begin{flushright}
  {\large HUTP-94/A006}
\end{flushright}

\vskip 0.4in

\begin{center}

{\LARGE Isospin Analysis of Two-body B Decays}
\vskip 10pt
{\LARGE  and Test of Factorization}

\vskip 0.8in

{\large Hitoshi Yamamoto}
\vskip 5pt
{\large\it Harvard University      \linebreak
 42 Oxford St., Cambridge, MA 02138, USA }

\end{center}

\vskip 0.5in

\begin{abstract}
It is shown that the existing data on two-body B decays, some of them only
upper limits, are precise enough to perform an isospin analysis to extract
the phase shifts due to final
state interaction. Unlike charm decays, no significant final
state interaction is observed in decays $B\to D\pi, D\rho$, and $D^*\pi$
supporting the factorization hypothesis in these decays.
{}From the isospin amplitudes obtained, we extract the ratio $a_2/a_1$, where
$a_1$ and $a_2$ are the coefficients in the factorized effective
Hamiltonian.
\end{abstract}

\vskip 0.25in

The idea of factorization has been used for evaluating non-leptonic weak
decays ever since Schwinger employed it to show that the $\Delta I = 3/2$ kaon
decay rate is consistent with the corresponding semileptonic decay
rate \cite{Schwinger,Feynman}. Even though the method was originally
thought to be useful only for order of magnitude estimations, it has been
extensively applied to heavy hadron decays \cite{Fak-Ste,AKKW,Bau-Ste,BSW}
with results varying from mixed to reasonable. Since factorization is
thus far virtually the only way to quantitatively calculate exclusive
non-leptonic
rates, it is important to test the validity of the hypothesis whenever
possible; in fact, it has been a subject of a number of studies
\cite{FacTest}.

We take the decay $\overline B^0 \to D^+\pi^-$ as an example
which can occur by the 4-fermion operator \cite{Bau-Ste}
\begin{equation}
   a_1(\overline d u)_\mu(\overline c b)^\mu +
   a_2(\overline c u)_\mu(\overline d b)^\mu
     \label{eq:effH}
\end{equation}
where the notation is $(\overline q q')_\mu\equiv
\overline q \gamma_\mu (1-\gamma_5) q'$ and $a_{1,2}$ are real coefficients
which are related to the Wilson coefficients $C_1$ and $C_2$ by
\begin{equation}
      a_1 = C_1 + \xi C_2 \qquad
      a_2 = C_2 + \xi C_1 .
    \label{eq:a12}
\end{equation}
The parameter $\xi$ is sometimes called
the color suppression factor and naively expected to be 1/3. At the mass
scale of $b$-quark, the leading-logarithm approximation gives \cite{LLA}
\begin{equation}
    C_1(m_b) = 1.11 \qquad C_2(m_b) = -0.26 .
\end{equation}

The basic idea
of factorization is that the pion is generated from vacuum by the current
operator $(\overline d u)_\mu$ and the transition $B\to D$ is caused by the
current operator $(\overline c b)^\mu$, and that they occur independently.
In terms of matrix element, it amounts to the fact that it can be written
in a factorized form:
\begin{equation}
   Amp(\overline B^0 \to D^+\pi^-) =
     {G_F \over \sqrt2} V^*_{ud}V_{cb}\,  a_1
        \langle \pi^- | (\overline d u)^\mu | 0 \rangle
     \langle D^+ | (\overline c b)_\mu | \overline B^0 \rangle .
     \label{eq:A+-}
\end{equation}
Such assumption of factorization
has been shown to be correct to the zeroth order in $1/N$
expansion \cite{Buras1/N}. Also, it has been argued intuitively that
factorization should hold for energetic tow-body decays
\cite{Fak-Ste,BjorkenFact} based on two observations: 1) when
the $u\overline d$ pair escapes the color field around the $b$ quark, it is
highly energetic ($\sim 2.5$ GeV) and, since it has to eventually
form a pion, the pair is collinear and close together with the total color
being zero. Thus, we expect that the pair will escape the color field
without much interaction (`color transparency'). 2) By the time the pion is
formed, it will be well outside the color field (again due to the high
energy of the pion); thus, there will be little
final state interaction (FSI) between the $D$ meson and the pion.
The same arguments had been used for $\rho$ and $\psi$
productions in hard scatterings \cite{Brodsky}, and recently put forward
further by Dugan and Grinstein in the framework of heavy quark effective
theory \cite{Dug-Gri}.

Since factorization assumes independence of the pion formation and
the $B$ to $D$ transition, FSI between them is
antithesis to factorization. Often, however, factorization and
FSI are combined to be compared with data assuming
that the factorization calculation correctly estimates the amplitude
`just before' FSI takes place \cite{BSW,BFMR}. This procedure
has worked reasonably well for charm decays, and we will
effectively employ it later
when we extract $a_2/a_1$.
It is worth keeping in mind, however, that such treatment is not justified in
the $1/N$ expansion \cite{Buras1/N} and also it is not well defined exactly
where factorization ends and FSI begins.

In the isospin analysis of the charm
decays $D\to K\pi$ \cite{BSW}, it is found that the FSI phase shift between
isospin 1/2 amplitude and isospin 3/2 amplitude is large ($\sim 77^\circ$).
Furthermore, after the effect of FSI is removed as described above, the
coefficients $a_1$ and $a_2$ are found to be nearly identical to the values
of Wilson coefficients $C_1$ and $C_2$ evaluated at the charm mass scale,
corresponding to
$\xi\sim0$. This has led to the rule of so-called `discarding 1/N terms'
\cite{Buras1/N}
and prompted further theoretical studies based on
QCD sum rules \cite{BlokSh}. A recent analysis \cite{CLEO}, however,
indicates that the situation is quite different for $B$ decays
giving the value of $\xi$ around 1/2 to 1/3.
In this study, we will see that the existing data leads us
to conclude that the FSI phase shifts are small for the $B$ decays
$B\to D\pi,D\rho$, and $D^*\pi$, and that the value of $\xi$ remains to be
1/2$\sim$1/3 even after the effect of FSI is taken out.

The Hamiltonian responsible for $B\to D\pi$ decays has isospin
$|1,-1 \rangle$, and this leads to the following isospin relations:
\begin{eqnarray}
 A^{+-} &=& \sqrt{1\over3}A_{3\over2} + \sqrt{2\over3}A_{1\over2}
          \nonumber \\
 A^{00} &=& \sqrt{2\over3}A_{3\over2} - \sqrt{1\over3}A_{1\over2}
             \label{eq:isorel} \\
 A^{0-} &=& \sqrt{3}A_{3\over2} \nonumber
\end{eqnarray}
where
$A^{+-}\equiv Amp(\overline B^0\to D^+\pi^-)$,
$A^{00}\equiv Amp(\overline B^0\to D^0\pi^0)$,
$A^{0-}\equiv Amp(B^-\to D^0\pi^-)$, and
$A_{3/2}$, $A_{1/2}$ are the isospin 3/2 and 1/2
amplitudes, respectively. There are three unknown parameters:
$|A_{3/2}|$, $|A_{1/2}|$, and
$\delta=\arg(A_{3/2}/A_{1/2})$. Since there are three measurements
of decay rates (namely, $|A_{+-}|^2$, $|A_{00}|^2$, and $|A_{0-}|^2$),
one can solve for the three unknowns:
\begin{eqnarray}
    |A_{3\over2}| &=& \sqrt{|A_{0-}|^2\over3}  \nonumber  \\
    |A_{1\over2}| &=& \sqrt{|A_{+-}|^2 + |A_{00}|^2 - {|A_{0-}|^2\over3}}
      \label{eq:anasol}  \\
    \cos\delta    &=&
       \frac{ |A_{+-}|^2 + |A_{00}|^2 - |A_{0-}|^2/3 }
        { \sqrt{{8\over3}|A_{0-}|^2
         (|A_{+-}|^2 + |A_{00}|^2 - |A_{0-}|^2/3)} } \nonumber
\end{eqnarray}
Note also that the isospin relations (\ref{eq:isorel})
can be expressed as a single triangle relation:
\begin{equation}
   A^{+-} + \sqrt2 A^{00} =  A^{0-} . \label{eq:isotri}
\end{equation}

The same relations hold for the decays $B\to D^*\pi$ and $B\to D\rho$. For
the decay $B\to D^*\rho$ the same relations hold separately for each
helicity amplitude. Since there is no a priori reason to believe that
the polarization is the same for $D^{*+}\rho^-$, $D^{*0}\rho^0$, and
$D^{*0}\rho^+$, and since there is not enough data to separate the helicity
amplitudes, we will not include $D^*\rho$ mode in this analysis.

Amplitudes calculated by factorization naturally satisfy the triangle isospin
relation (\ref{eq:isotri}). This can be seen from the expression
(\ref{eq:A+-}) and corresponding factorized forms for $A_{+-}$ and $A_{00}$:
\begin{eqnarray}
   A^{00} &=&
     {G_F \over \sqrt2} V^*_{ud}V_{cb}
        a_2\langle D^0 | (\overline c u)^\mu | 0 \rangle
     \langle \pi^0 | (\overline d b)_\mu | \overline B^0 \rangle
         \nonumber \\
   A^{0-} &=&
     {G_F \over \sqrt2} V^*_{ud}V_{cb}
       \Big[ a_1 \langle \pi^- | (\overline u d)^\mu | 0 \rangle
     \langle D^0 | (\overline c b)_\mu | \overline B^- \rangle
         \nonumber \\
          &&\qquad\qquad
       + a_2 \langle D^0 | (\overline c u)^\mu | 0 \rangle
     \langle \pi^- | (\overline d b)_\mu | \overline B^- \rangle \Big]
         \nonumber
\end{eqnarray}
and noting that (from isospin symmetry)
\begin{eqnarray}
     \langle D^0 | (\overline c b)_\mu | \overline B^- \rangle &=&
     \langle D^+ | (\overline c b)_\mu | \overline B^0 \rangle
              \nonumber \\
     \sqrt2\langle \pi^0 | (\overline d b)_\mu | \overline B^0 \rangle &=&
     \langle \pi^- | (\overline d b)_\mu | \overline B^- \rangle
              \nonumber  .
\end{eqnarray}
In fact, the isospin amplitudes are explicitly given by
\begin{eqnarray}
    A_{3\over2} &=&
     {G_F \over \sqrt2} V^*_{ud}V_{cb}\,
      {1\over\sqrt3}(M_1 + M_2) \label{eq:faciso} \\
    A_{1\over2} &=&
     {G_F \over \sqrt2} V^*_{ud}V_{cb}\,
      \sqrt{3\over2} \Big( {2\over3}M_1 - {1\over3}M_2 \Big)
           \nonumber
\end{eqnarray}
or
\begin{equation}
   {A_{3/2}\over A_{1/2}} =
        \sqrt2\,{ 1 + M_2/M_1 \over 2 - M_2/M_1}
       \label{eq:isoratio}
\end{equation}
where
\begin{eqnarray}
   M_1 &\equiv& a_1
     \langle \pi^- | (\overline d u)_\mu | 0 \rangle
     \langle D^+ | (\overline c b)^\mu | \overline B^0 \rangle
     \label{eq:M12} \\
   M_2 &\equiv& a_2
     \langle D^0 | (\overline c u)_\mu | 0 \rangle
     \langle \pi^- | (\overline d b)^\mu | B^- \rangle
     \nonumber
\end{eqnarray}
As before, the same relations (\ref{eq:faciso}-\ref{eq:M12}) hold for
decays $D^*\pi$, $D\rho$, and each helicity state of $D^*\rho$.

If factorization is assumed, the three amplitudes
$A_{+-}, A_{00}$, and $A_{0-}$ are relatively real, and thus
the triangle (\ref{eq:isotri}) reduces to a line. Therefore, if the
measured decay
rates are exactly as expected from factorization as prescribed above, then
the isospin analysis is guaranteed to give $\delta = 0$. Actual
measurements, however, are always associated with errors, and the main
point of this article is in showing that meaningful isospin analyses can be
performed even though only upper limits are available for some of the
decay modes.

The isospin amplitudes cannot be uniquely given by factorization since it
depends on decay constants and form factors through (\ref{eq:M12}). If we
use the model of Bauer, Stech and Wirbel \cite{BSW,HFreview}
together with $f_\pi = 132$ MeV and $f_D = 220$ MeV, we obtain
\begin{equation}
     {M_2 \over M_1} = \left\{
           \begin{array}{ll}
               1.23\, a_2/a_1 &\quad (D\pi)  \\
               1.30\, a_2/a_1 &\quad (D^*\pi) \\
               0.66\, a_2/a_1 &\quad (D\rho)
           \end{array}
                       \right.
     \label{eq:M2-M1}
\end{equation}
where the $a_1, a_2$ are the coefficients appearing in the effective
Hamiltonian (\ref{eq:effH}). This can be substituted in (\ref{eq:isoratio})
to obtain the expected ratio of isospin amplitudes, or if the ratio
$A_{3/2}/A_{1/2}$ is known, $a_2/a_1$ can be extracted from
\begin{equation}
    {M_2\over M_1} = 2\,{A_{3/2}/A_{1/2} - 1/\sqrt2 \over
                         A_{3/2}/A_{1/2} +   \sqrt2 } .
     \label{eq:M2-M1b}
\end{equation}

The table 1. shows the current available measurements for the relevant
decay modes \cite{CLEO}.
\begin{table}
\begin{center}
\begin{tabular}{|cc|cc|cc|}
\hline
\hline
     $\overline B^0$ mode & (\%) &
     $\overline B^0$ mode & (\%) &
     $B^-$ mode & (\%) \\
\hline
  $D^+\pi^-$      & $0.29\pm0.04$       &
  $D^0\pi^0$      & $<0.035$            &
  $D^0\pi^-$      & $0.55\pm0.04$            \\
                  & $\pm0.03\pm0.05$
               &&&& $\pm0.03\pm0.02$         \\

  $D^{*+}\pi^-$   & $0.26\pm0.03$       &
  $D^{*0}\pi^0$   & $<0.072$            &
  $D^{*0}\pi^-$   & $0.49\pm0.07$            \\
                  & $\pm0.03\pm0.01$
               &&&& $\pm0.06\pm0.03$         \\

  $D^+\rho^-$     & $0.81\pm0.11$       &
  $D^0\rho^0$     & $<0.042$            &
  $D^0\rho^-$     & $1.35\pm0.12$            \\
                  & $\pm0.12\pm0.13$
               &&&& $\pm0.12\pm0.04$         \\
\hline
\end{tabular}
\end{center}
\caption{Branching ratios measured by CLEO.
The first error is statistical,
the second error is systematic, and the third error is due to uncertainties
in $D$ branching ratios. The upper limits are 90\%\ confidence levels.}
\label{tb:CLEO}
\end{table}
We will take the statistical errors only, and assume
that the life times of $\overline B^0$ and $B^-$ are the same:
\begin{equation}
    \tau(\overline B^0) = \tau(B^-) = 1.18 {\rm ps}. \nonumber
\end{equation}
The upper limits are converted to a gaussian distribution centered at zero
by setting the r.m.s. of the gaussian to (upper limit)/1.64.

Table \ref{tb:result} shows the solution for $|A_{3/2}|, |A_{1/2}|$ and
$\cos\delta$ using the formulae
(\ref{eq:anasol}). For each mode, $\cos\delta$ is consistent with unity
indicating that there is no phase shifts due to final state interaction.
In this analytical method, however, the range of $\cos\delta$ is not
constrained to within $\pm1$. In order to take the constraint into account
properly, we will use the maximum likelihood method.
The likelihood
function for $|A_{3/2}|$, $|A_{1/2}|$, and $\cos\delta$ is given by
\begin{equation}
    L = N \prod_{i=+-,00,0-} {1\over\sqrt{2\pi}\sigma_i}
          \exp\left({(\Gamma_i - \Gamma^0_i)^2\over 2\sigma_i^2}\right)
\end{equation}
where $\Gamma_i$ and $\sigma_i$ are the measured decay rate and its error,
$\Gamma^0_i$ is the decay rate calculated from the amplitude $A^i$ given
by (\ref{eq:isorel}) according to the standard formula
\begin{equation}
     \Gamma^0_i = {p\over 8\pi M_B^2} |A^i|^2 \nonumber ,
\end{equation}
and the normalization factor $N$ is added to make the integral
over the allowed region of $|A_{3/2}|$, $|A_{1/2}|$, and $\cos\delta$
to be unity.
\begin{table}
\begin{center}
\begin{tabular}{|cc|c|c|c|}
\hline
\hline
    \multicolumn{2}{|c|}{} & $D\pi$ & $D^*\pi$ & $D\rho$ \\
\hline
    analytical & $|A_{3/2}|$ ($10^{-5}$ GeV)
                                  & $0.556\pm0.021$
                                  & $0.533\pm0.038$
                                  & $0.886\pm0.040$  \\
    solution   & $|A_{1/2}|$ ($10^{-5}$ GeV)
                                  & $0.425\pm0.095$
                                  & $0.411\pm0.127$
                                  & $0.792\pm0.133$  \\
               & $\cos\delta$     & $1.20\pm0.28$
                                  & $1.19\pm0.61$
                                  & $1.11\pm0.13$   \\
 \hline
    maximum    & $|A_{3/2}|$ ($10^{-5}$ GeV)
                                  & $0.550\pm0.020$
                                  & $0.527\pm0.037$
                                  & $0.862\pm0.037$  \\
    likelihood & $|A_{1/2}|$ ($10^{-5}$ GeV)
                                  & $0.503\pm0.057$
                                  & $0.494\pm0.065$
                                  & $0.907\pm0.084$  \\
               & $\cos\delta^*$   & $>0.82$
                                  & $>0.57$
                                  & $>0.92$        \\
 \hline
    \multicolumn{2}{|c|}{$|A_{3/2}/A_{1/2}|$}
               & $1.09\pm0.13$ & $1.07\pm0.16$ & $0.95\pm0.10$ \\
    \multicolumn{2}{|c|}{$a_2/a_1$}
               & $0.25\pm0.07$ & $0.22\pm0.08$ & $0.31\pm0.11$ \\
\hline
\end{tabular}
\end{center}
    $^*$ In all cases the most likely value for $\cos\delta$ is unity.
         The lower limits are at 90\%\ confidence level.
\caption{Analytical solutions
         and results of the maximum likelihood fit
         for the isospin amplitudes and
         their relative phase angle. Also given are the ratio of the
         isospin amplitudes and $a_2/a_1$ derived therefrom (using the
         result of the maximum likelihood fit).}
\label{tb:result}
\end{table}
Result of the fit is also shown in Table \ref{tb:result}. Because of the
constraint on $\cos\delta$, the errors are generally better than those
of the analytical solutions. For each mode, the most likely value for
$\cos\delta$ was unity. Figure \ref{fg:1} show the 1,2 and 3 sigma contours
for $|A_{3/2}|$ vs $|A_{1/2}|$ and $|A_{3/2}|$ vs $\cos\delta$.
It is seen that the parameters are not strongly correlated.
\begin{figure}
  \vspace{5.5in}
  \caption{One, two and three sigma contours for the 2-dimentional
   plots of the likelihood function: $|A_{3/2}|$ vs $|A_{1/2}|$ (a) and
   $|A_{3/2}|$ vs $\cos\delta$ (b).}
  \label{fg:1}
\end{figure}

Also shown in Table \ref{tb:result} are the ratio of isospin amplitudes
$|A_{3/2}/A_{1/2}|$ and $a_2/a_1$ extracted using equations
(\ref{eq:M2-M1}-\ref{eq:M2-M1b}).
For each mode, the ratio $a_2/a_1$ is positive which is a consequence of
$|A_{1/2}|<\sqrt2|A_{3/2}|$.
Averaging over the three modes, we obtain
\begin{equation}
     {a_2\over a_1} = + 0.25\pm0.05  \nonumber
\end{equation}
or from (\ref{eq:a12}) the corresponding color suppression factor $\xi$ is
\begin{equation}
     \xi = 0.45\pm0.04   \nonumber
\end{equation}
which is consistent with the analysis of Ref \cite{CLEO} where the decay
rates were fit to the model by Bauer, Stech and Wirbel without taking out
the final state interaction.
This, however, cannot be considered to be an
independent confirmation of the positive value of $a_2/a_1$ since the two
analyses are highly correlated.

In summary, we have performed an isospin analysis on two-body B decays and
found that the phase shifts by final state interaction are small in stark
contrast to the case of charm decays.
By fitting to the obtained isospin amplitudes, we have also
seen that the effect of removing
the final state interaction does not alter the observation that the
ratio $a_2/a_1$ is positive.

\begin{center}
{\bf Acknowledgement}
\end{center}

The author would like to thank Mike Dugan, Giulia Ricciardi, Thorsten
Ohl, and Mitch Golden for fruitful discussions.
This research was supported primarily by DOE
grant DEFG-029-1ER-40-654, DOE Outstanding Junior Investigator program,
Milton Fund of Harvard Medical School,
and in part by NSF grant PHY-92-18167.

\pagebreak

\end{document}